\documentclass[twocolumn,aps,prl,amsmath,showpacs,nofootinbib]{revtex4-1}
\pdfoutput=1
\usepackage{graphicx,bm,amssymb,amsmath}
\usepackage{color}

\usepackage{float}

\usepackage{hyperref}
\def\be{\begin{equation}}
\def\ee{\end{equation}}

\usepackage{hyperref}

\begin{document}
\title{Theory of phonon spectrum in host-guest crystalline solids with avoided crossing}
\author{M. Baggioli$^{1}$, B. Cui$^{2}$ and   A. Zaccone$^{3,4,5}$}

\affiliation{${}^1$Instituto de Fisica Teorica UAM/CSIC, c/Nicolas Cabrera 13-15,
Universidad Autonoma de Madrid, Cantoblanco, 28049 Madrid, Spain.}
\affiliation{${}^2$Cavendish Laboratory, University of Cambridge, JJ Thomson
Avenue, CB30HE Cambridge, U.K.}
\affiliation{${}^3$Department of Physics "A. Pontremoli'', University of Milan, via Celoria 16, 20133 Milan, Italy}
\affiliation{${}^4$Department of Chemical Engineering and Biotechnology,
University of Cambridge, Philippa Fawcett Drive, CB30AS Cambridge, U.K.}
\begin{abstract}
\noindent 
We develop an analytical model to describe the phonon dispersion relations of host-guest lattices with heavy guest atoms (rattlers). Crucially, the model also accounts for phonon damping arising from anharmonicity. The spectrum of low energy states contains acoustic-like and (soft) optical-like modes, which display the typical avoided crossing, and which can be derived analytically by considering the dynamical coupling between host lattice and guest rattlers. Inclusion of viscous anharmonic damping in the model allows us, for the first time, to compute the vibrational density of states (VDOS) and the specific heat, unveiling the presence of a boson peak (BP) linked to an anharmonicity-smeared van Hove singularity. Upon increasing the coupling strength between the host and the guest dynamics, and by decreasing the energy of the soft optical modes, the BP anomaly becomes stronger and it moves towards lower frequencies. Moreover, we find a robust linear correlation between the BP frequency and the energy of the soft optical-like modes. This framework provides a useful model for tuning the thermal properties of host-guest lattices by controlling the VDOS, which is crucial for optimizing thermal conductivity and hence the energy conversion efficiency in these materials.
\end{abstract}
{
\maketitle
}

In glasses, contrarily to crystalline structures with long-range order, standard propagating phononic modes with ballistic dispersion relation $\omega\,=\,v_{L,T}\,q$, where $L,T$ refer to longitudinal and transverse modes \cite{phillips1987two,allen1999diffusons,baggioli2019hydrodynamics} are not the only or the dominant vibrational excitations. Rather, the breakdown of continuum elasticity at sufficiently low length scales generates a proliferation of quasi-localized modes which are characterized by diffusive-like propagation due to scattering~\cite{allen1999diffusons}. Both dissipation-less (''harmonic") scattering due to static disorder as well as scattering due to anharmonicity contribute to an excess of vibrational modes which appears as a peak (the boson peak) in the vibrational density of states (VDOS) when normalized by the Debye $\sim \omega^{2}$ law of ballistic phonons.
This observation is the fundamental reason behind the anomalies observed experimentally in the VDOS, the specific heat and the thermal conductivity of amorphous and strongly disordered systems \cite{nakayama2002boson,PhysRevB.4.2029}. 

\begin{figure}
\includegraphics[width=\linewidth]{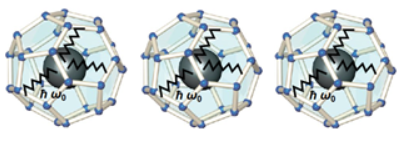}
\caption{A pictorial representation of guest-host interplay associated with the presence of quasi-localized rattler modes. The host-guest lattice is approximated by a finite concentration of caged guest atoms or "defects'' (typically, heavier atoms compared to the host lattice atoms). Each defect consists of a caged guest atom with a single localized soft mode with energy $E=\hbar \omega_0 \ll \hbar \omega_D$, where 
$\omega_0$ is the energy of the optical-like modes and $\omega_{D}$ is the Debye frequency. Upon increasing the concentration $c$ of the caged guest atoms, the glassy phonon features become more and more pronounced.}
\label{fig0}
\end{figure}

Recently, the observation of glassy, or more precisely glassy-like, anomalies has been extended to systems with minimal or orientational disorder and simple crystalline structures \cite{PhysRevLett.119.215506,doi:10.1021/acs.jpcb.5b04240,PhysRevLett.112.025502,PhysRevB.99.024301,PhysRevB.94.224204}. A possible theoretical explanation has been proposed for these systems~\cite{BaggioliPRL}, in terms of the interplay of elasticity and ballistic propagation with damping and effective viscosity. Additionally, glassy features (referred as "phonon glass'') have been observed in thermoelectric host-guest materials \cite{rowe1995crc,rowe2018thermoelectrics,RevModPhys.86.669,PhysRevB.77.235119,doi:10.1063/1.4955398}, such as clathrates, tetrahedrites and skutterudites with guest inclusions (rattlers) which display interesting thermal transport features and which could provide important technological applications \cite{SALES1997284}. 

Perhaps the most prominent quality of thermoelectrics is their ability to conduct electricity efficiently, like a crystalline solid, and at the same time to conduct heat poorly, like a glass, i.e. the "phonon glass-electron crystal'' paradigm \cite{Nakayama_2008,doi:10.7566/JPSJ.84.103601}. Many of these systems have a perfectly ordered crystalline (host) lattice but they contain caged (typically heavier) atoms referred to as "rattlers'' (see Fig.\ref{fig0}), which give rise to quasi-localized vibrational modes  \cite{PhysRevLett.114.095501,PhysRevLett.120.105901,doi:10.1002/adma.201706230,PhysRevLett.106.015501}. The interactions between the guest rattlers and the host lattice modes might be crucial to understand and control the thermal conductivity of these crystalline materials, although a theory of this effect is still lacking. Moreover, the presence of the rattlers produce the avoided crossing phenomenon \cite{doi:10.1119/1.3471177}, which is typical of the host-guest materials \cite{doi:10.1063/1.475218,Christensen2008}.\\

In this work we focus on a specific framework, known as soft-mode dynamics theory (SMD) \cite{1981SSCom..37..975K,PhysRevB.46.2798,Klinger2008}, and in particular on the simple theoretical model proposed by Klinger and Kosevich (KK) in ~\cite{KKmodel}. In the original KK framework, the model consists of the usual elastodynamic equation for the displacement field of an elastic solid with an extra term given by the dynamic (mutual) coupling to the coordinate of a defect particle (the guest atom). The dynamics of the latter is governed by a Newton's equation in a harmonic force-field with likewise an extra term due to the coupling with the elastic embedding solid.

\begin{figure}[H]
\centering
\includegraphics[width=0.8\linewidth]{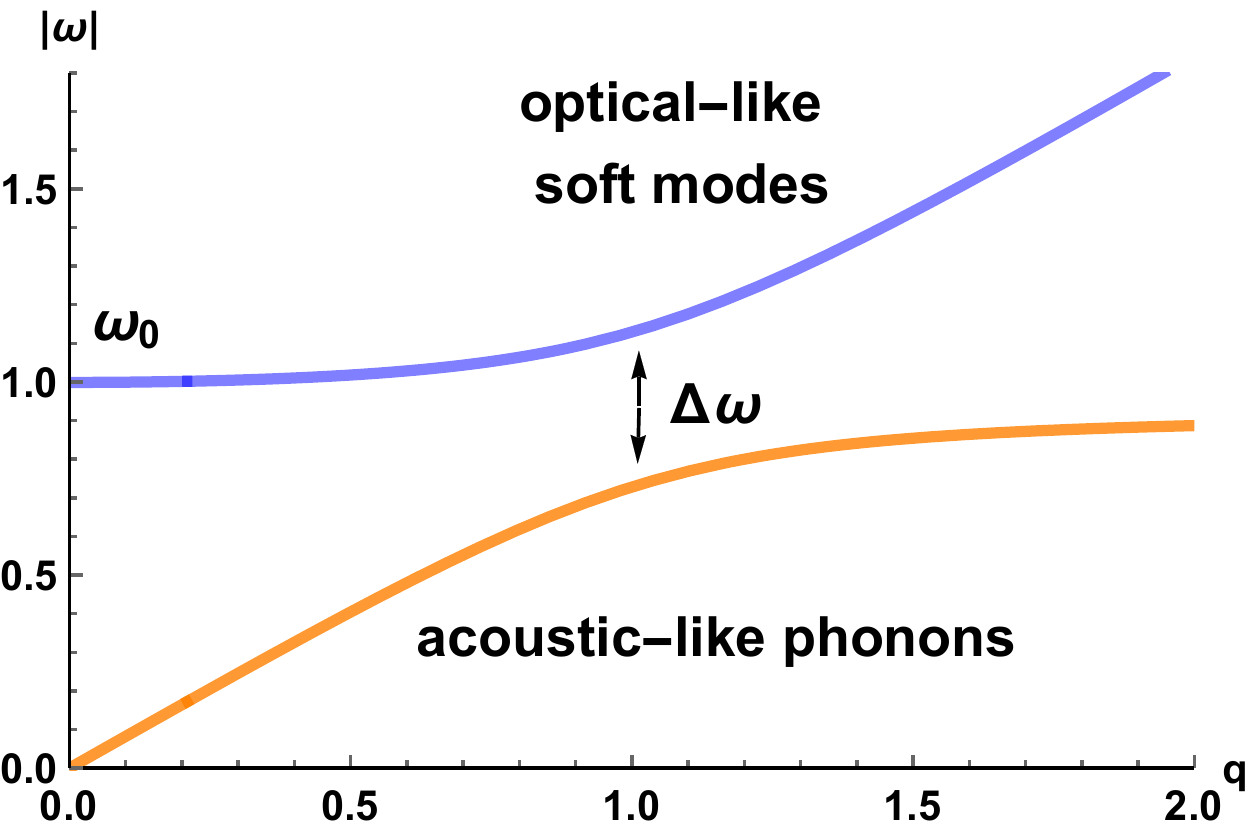}
\caption{(Color online) The two branches of low energy modes present in the KK model of \cite{KKmodel}. The avoided crossing is determined by an energy separation $\Delta \omega$ which is physically controlled by the concentration of defects $c$.}
\label{disp}
\end{figure}

The main result of the KK theory is a polariton-like phonon spectrum with the coexistence of acoustic-like and soft optical-like modes separated by a characteristic avoided-crossing feature, due to dynamical coupling of the defects to the elastic lattice matrix. The two branches display a distinctive avoided crossing feature. Furthermore they provide a close approximation for the phononics of thermoelectric materials such as e.g. $Ba_{8}Ga_{16}Ge_{30}$~\cite{Christensen2008} as well as clathrate hydrates~\cite{doi:10.1063/1.475218} and tetrahedrites~\cite{doi:10.1002/adma.201706230}. A representation is provided in Fig.\ref{disp}. Moreover, the typical energy of the gapped soft optical modes $\omega_0$ has been claimed to be related to the Ioffe-Regel energy scale $\omega_{IR}$, which is connected with the frequency $\omega_{BP}$ of the so-called boson peak anomaly (excess of Debye's law $\sim \omega^{2}$) measured in the vibrational spectrum of glasses \cite{KKmodel}.

However, the KK model has remained rather limited in its applicability for predictions of thermal properties of materials, because in its previous formulation it cannot provide access to the vibrational density of states (VDOS). The latter is the key quantity which enters the integrals that yield the specific heat and the thermal conductivity of a material. Furthermore, also the original speculation by Klinger and Kosevich that the soft-mode coupled dynamics in the KK model could lead to a boson peak in the VDOS, and its relation to a Ioffe-Regel crossover, have not been verified.

In this paper we provide a working answer to all these questions, by extending the KK model to realistic crystalline lattices where both wave propagation through the lattice as well as the rattler motion are damped by anharmonicity. This extension allows us, for the first time, to evaluate the VDOS analytically for host-guest lattices. The calculation reveals the presence of a boson peak in the VDOS, at a frequency which is close to the frequency of the van Hove peak occurring as the group velocity $d\omega/dk=0$. This effect results in a peak in the specific heat as well. Such analytical model, which gives access to the VDOS as  a function of key structural parameters such as defect density $c$ and the dynamic coupling between defect and matrix, may play a crucial role for the understanding and design of host-guest materials where thermal conductivity is controlled by the features of the VDOS.

\begin{figure}[h]
\centering
\includegraphics[width=0.49 \linewidth]{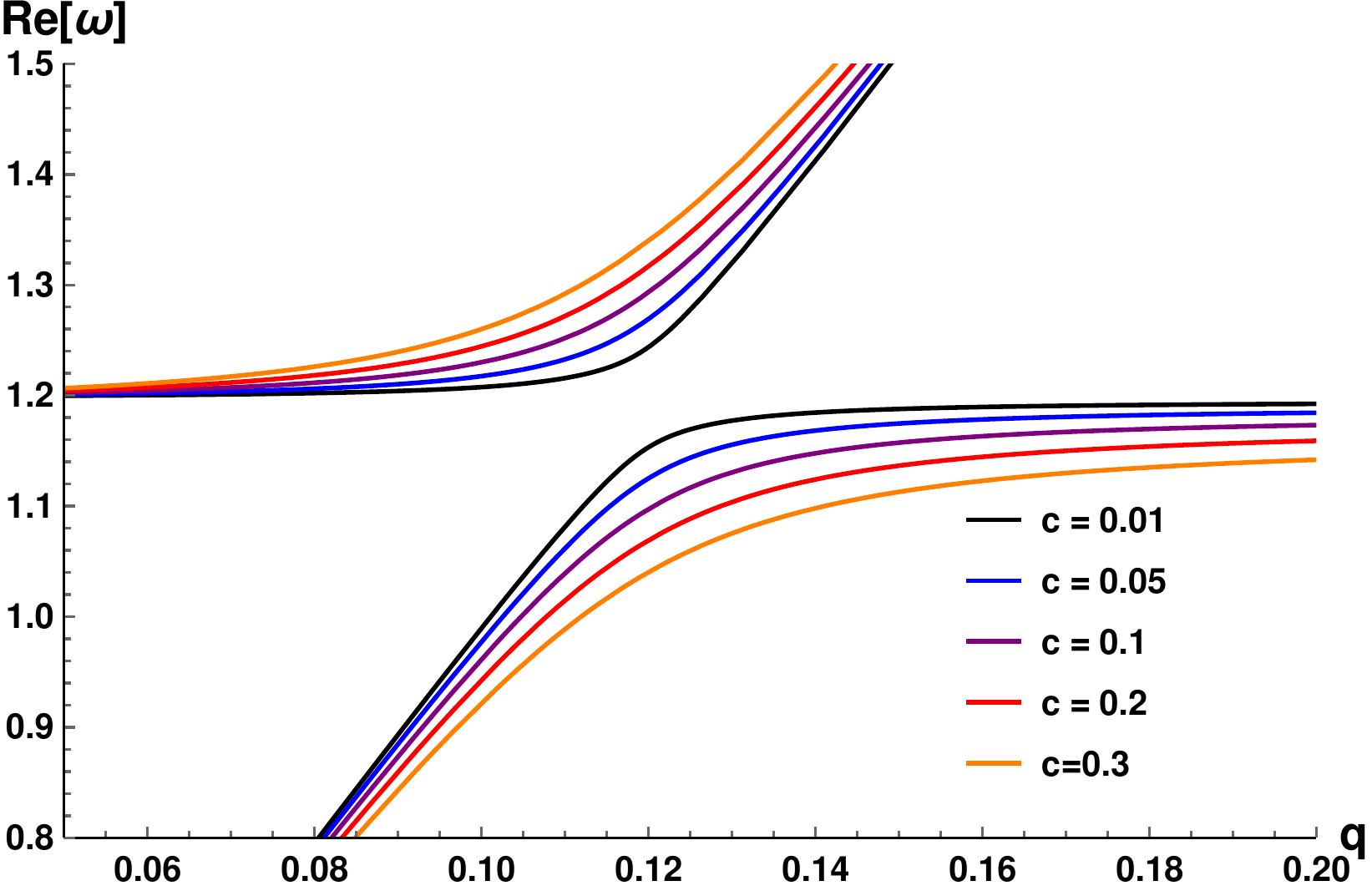}
\includegraphics[width=0.49 \linewidth]{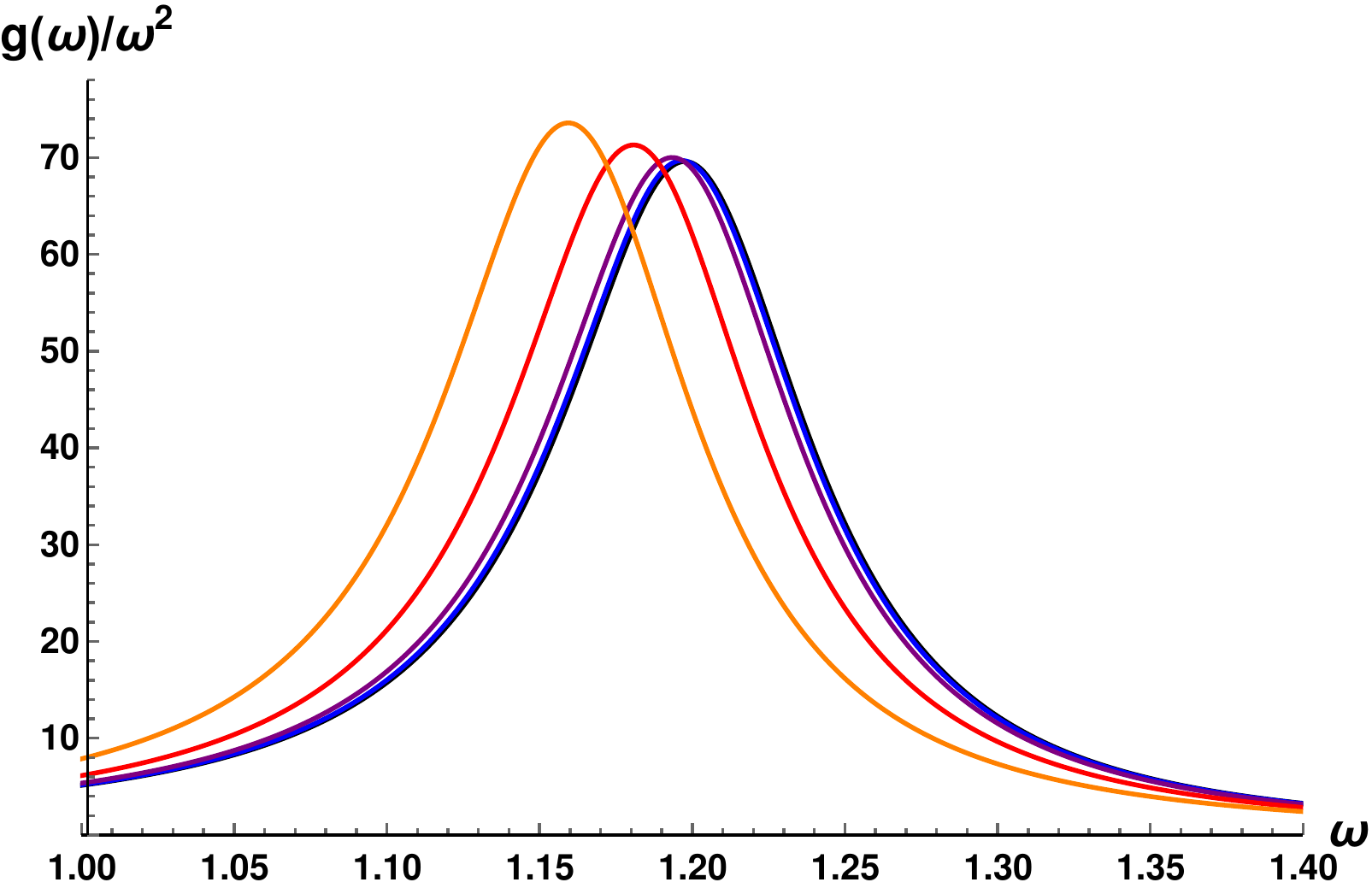}
\vspace{0.2cm}
\includegraphics[width=0.51 \linewidth]{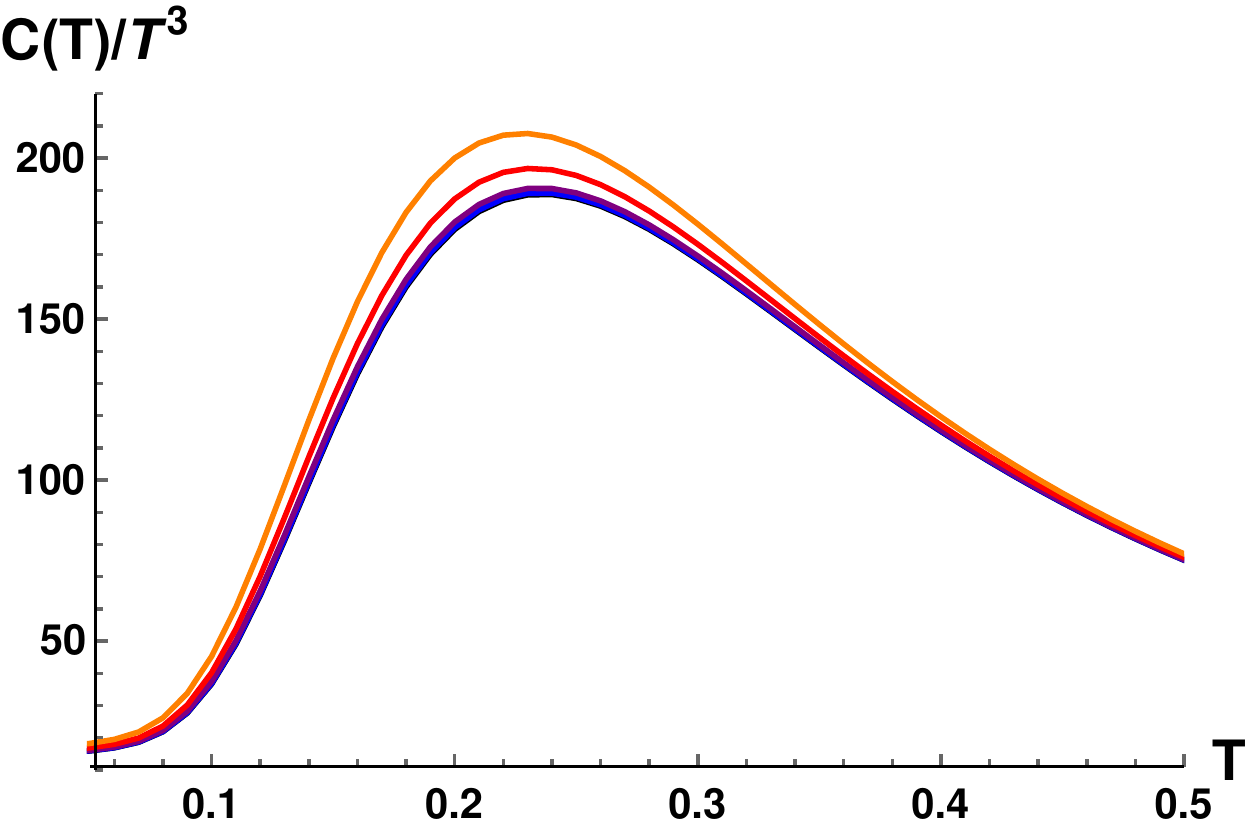}
\caption{(Color online) \textbf{Left: }The dispersion relation of the two-branches of modes in the extended KK model with damping (Eq.\eqref{uno}). The different colors correspond to different values of the density of defects, encoded in the parameter $c$, see legend. In all calculations the damping coefficients are $\gamma_1=1$, $\gamma_2=0.1$. \textbf{Right: } The corresponding normalized VDOS. \textbf{Bottom: } The normalized specific heat.}
\label{panel1}
\end{figure}

Let us briefly summarize the setup used for our derivations; for more details see  \cite{Supplementary}. In particular, we generalize the KK model \citep{KKmodel} by introducing finite anharmonic damping contributions for both the dynamics of the host lattice matrix (which becomes effectively a visceolastic medium) and the dynamics of the guest defect atom. The resulting equations read as~\cite{KKmodel}
\begin{align}\label{uno}
\begin{cases}
\rho\,\frac{\partial^2\mathbf{u}}{\partial^2t}\,\simeq\,\rho \,s_0^2\,\vartriangle\mathbf{u}+\frac{c\,\beta\,\partial x}{\partial\mathbf{R}}+\gamma_1\frac{\partial\bigtriangleup\mathbf{u}}{\partial t}\\
\mu\frac{\partial^2x}{\partial t^2}\simeq-\mu\,\omega_0^2x-\beta\,\epsilon(\mathbf{R})+\gamma_2\frac{\partial x}{\partial t}
\end{cases}
\end{align}
where $\rho$ is the density of the embedding viscoelastic matrix and $\mu$ is an effective mass parameter of the "soft mode'' guest atom. Furthermore, we defined the elastic displacement vector $\mathbf{u}$, the scalar strain parameter $\epsilon \equiv \text{div}\, \mathbf{u}$, the location of the defect site $\mathbf{R}$ and the soft mode dynamical (scalar) coordinate $x$, such that $U=\frac{1}{2}\mu\omega_{0}x^{2}$ is the potential energy of the soft mode, while $U_{int}=\beta\epsilon x$ is the coupling energy between the soft mode and the host matrix. The coupling strength is controlled by the parameter $\beta$ while the density of defects is given by $c$. The bare speed of sound is indicated with $s_0$ and the natural oscillation frequency of the soft mode with $\omega_0$. 

With respect to the KK model, we introduced a significant novelty in Eq.\eqref{uno} by adding dissipative coefficients $\gamma_{1,2}$ which determine the anharmonic damping of the acoustic-like phonons and of the soft modes. Notice that $\gamma_1$, the damping of the acoustic-like phonons, is modelled by adding a dissipative (viscous) term to the overall stress (as done in ~\cite{Lubensky} p.366), which leads to a diffusive-like dependence of the damping $\gamma_1 \sim q^{2}$ on the wavevector $q$. After some standard manipulations which involve solving the secular determinant, and by going to Fourier space, the key equation describing the vibrational modes of the system is  obtained as~\cite{Supplementary}
\begin{align}\label{modes}
&(\omega^2-s_0^2\,q^2+i\,\omega\,\frac{\gamma_1 \,q^2}{\rho})(\omega^2-\omega_0^2+i\,\omega\,\frac{\gamma_2}{\mu})=cQ^2\,s_0^2\,q^2
\end{align}
where we defined $cQ^2 \equiv c\,\beta^2\,/\rho\mu s_0^2\,<\,\omega_0^2$. An example of the polariton-like spectrum $\omega(q)$ with anharmonic damping computed using the above equation is shown in Fig.\ref{disp}.
In the regime of high defects concentration $c$, the interaction between the two types of modes becomes strong near the (avoided) intersection of the two branches, $\omega \sim \omega_0$, thus producing an avoided-crossing behavior. Notice that the energy separation $\Delta \omega$ is controlled by the strength of the interaction between the modes, $\beta$, and by the defects density $c$. More precisely, neglecting the subleading contributions coming from the damping coefficients, the energy separation between the two branches roughly reads as $\Delta \omega \sim \omega_0-\sqrt{\omega_0^2-cQ^2}$.

The main question we want to address here is how the above features affect the vibrational density of states (VDOS) and the specific heat of the system. We compute the VDOS using the standard formula
\begin{align}
g(\omega)\,=\,-\,\frac{2 \omega}{ \pi \,q_{D}^{3}}\,\int_0^{q_D}\,\text{Im}\,\mathcal{G}(q,\omega)\,q^2dq \label{form}
\end{align}
in terms of the Green's function $\mathcal{G}(q,\omega)$, which is derived from the Plemelj identity and was already derived and used in similar context in \cite{BaggioliPRL,baggioli2018soft,baggioli2019hydrodynamics}.
For the explicit form of the Green's function derived in full detail see Supplemental Material available at ~\cite{Supplementary}. Upon implementing it in Eq. \eqref{form}, we obtain the final semi-analytical expression for the reduced VDOS of the host-guest system lattice,
\begin{widetext}
\begin{equation}
\frac{g(\omega)}{\omega^2}\,=\,\frac{2}{\pi q_{D}^{3}}\,\int_0^{q_D}\frac{-\left[\left(\frac{\gamma_1q^2}{\rho}+\frac{\gamma_2}{\mu}\right)\omega^2-(\frac{\omega_0^2\gamma_1q^2}{\rho}+\frac{s_0^2q^2\gamma_2}{\mu})\right]q^2}{\left[\omega^4-\left(s_0^2q^2+\omega_0^2+\frac{\gamma_1q^2\gamma_2}{\rho\mu}\right)\omega^2+s_0^2q^2(\omega_0^2-cQ^2)\right]^2+\left[\left(\frac{\gamma_1q^2}{\rho}+\frac{\gamma_2}{\mu}\right)\omega^3-(\frac{\omega_0^2\gamma_1q^2}{\rho}+\frac{s_0^2q^2\gamma_2}{\mu})\omega\right]^2}\,dq\label{dos}
\end{equation}
\end{widetext}
where the poles of the integrand correspond to the roots of Eq. \eqref{modes}.

We can then derive the specific heat of the system by performing the standard integral~\cite{Khomskii}
\begin{equation}
C(T)\,=\,k_B\,\int_0^\infty \,\left(\frac{\hbar\omega}{2\,k_B\,T}\right)^2\,\sinh \left(\frac{\hbar\omega}{2\,k_B\,T}\right)^{-2}\,g(\omega)\,d\omega. \label{ct}
\end{equation}
\begin{figure}[h]
\centering
\includegraphics[width=0.49 \linewidth]{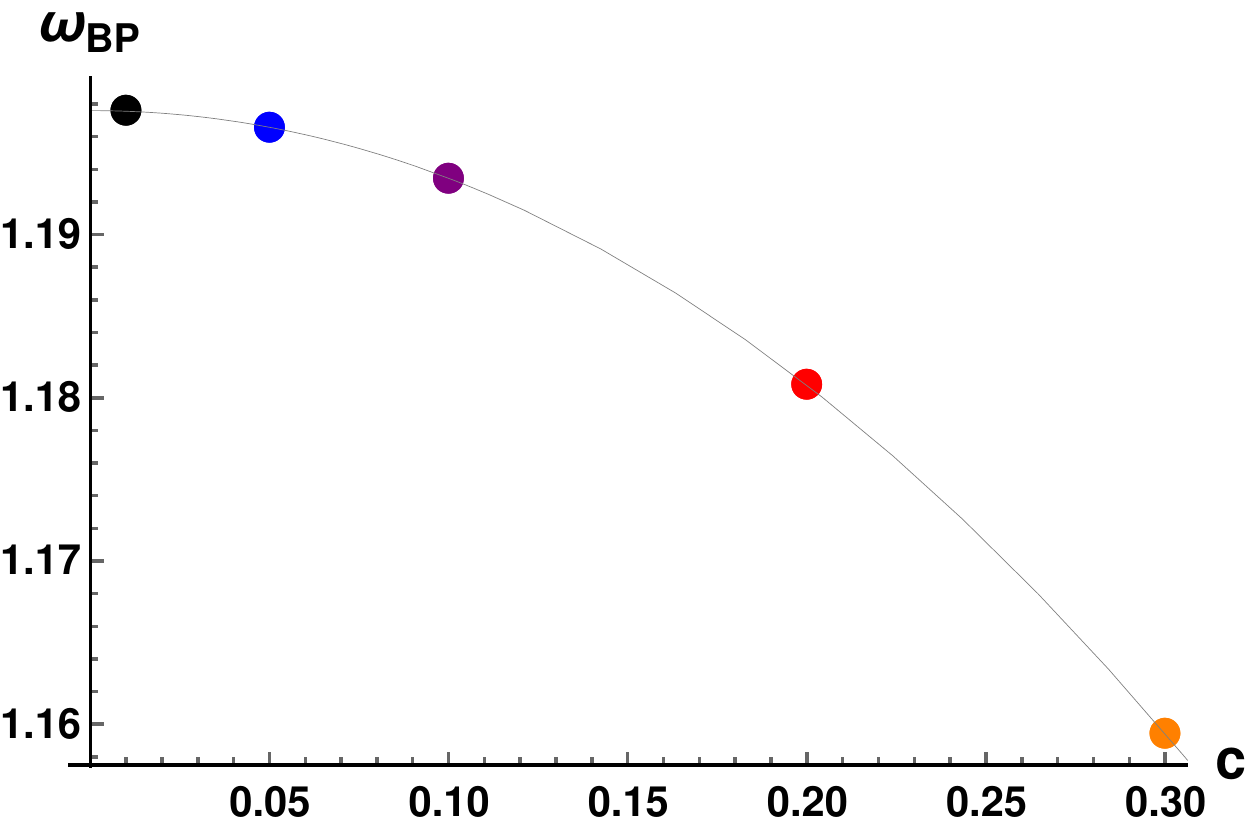}
\includegraphics[width=0.49 \linewidth]{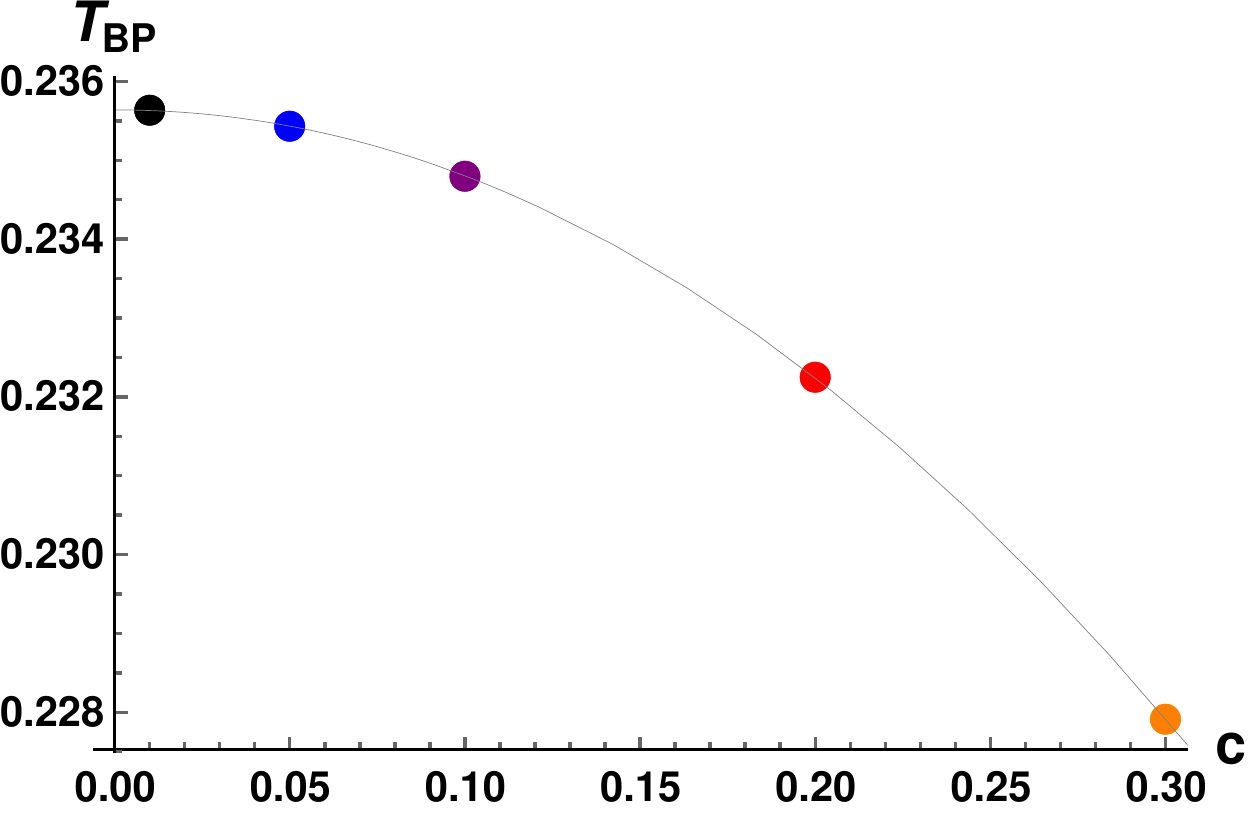}
\caption{(Color online) The evolution of the boson peak frequency $\omega_{BP}$ and boson peak temperature $T_{BP}$ as a function of the defect guest atom density $c$. The solid lines are empirical fits to power-law functions with power exponents 2.02 and 2.02 for $\omega_{BP}$ and $T_{BP}$, respectively.}
\label{panel2}
\end{figure}\\

Using formulae Eq.\eqref{dos} and Eq.\eqref{ct}, we are now ready to study in detail the features of the generalized KK model.
First we analyze the behavior of the system by varying the parameter $c$, which controls the density of the guest atoms while keeping fixed the characteristic energy of the soft mode, $\omega_0$. Upon increasing the density of defects we expect the interactions between the two types of modes to increase, the avoided crossing to be more pronounced and the glassy features to be more evident. The spectrum can be observed in the left panel of Fig.\ref{panel1}, for some benchmark values of $c$. As already mentioned, by increasing this parameter, the avoided crossing becomes stronger and the energy separation $\Delta \omega$ increases.

The reduced VDOS is shown in the right panel of Fig.\ref{panel1}. Increasing the density of defects $c$, the boson peak becomes more pronounced and it shifts towards low energies. The same phenomenon occurs in the specific heat of the system normalized by the Debye $\sim T^{3}$ contribution (see the bottom panel of Fig.\ref{panel1}). The position of the boson peak frequency and boson peak temperature follow a power-law scaling with the density of defects $c$ which is apparent in Fig.\ref{panel2}.\\

\begin{figure}[h]
\centering
\includegraphics[width=0.49 \linewidth]{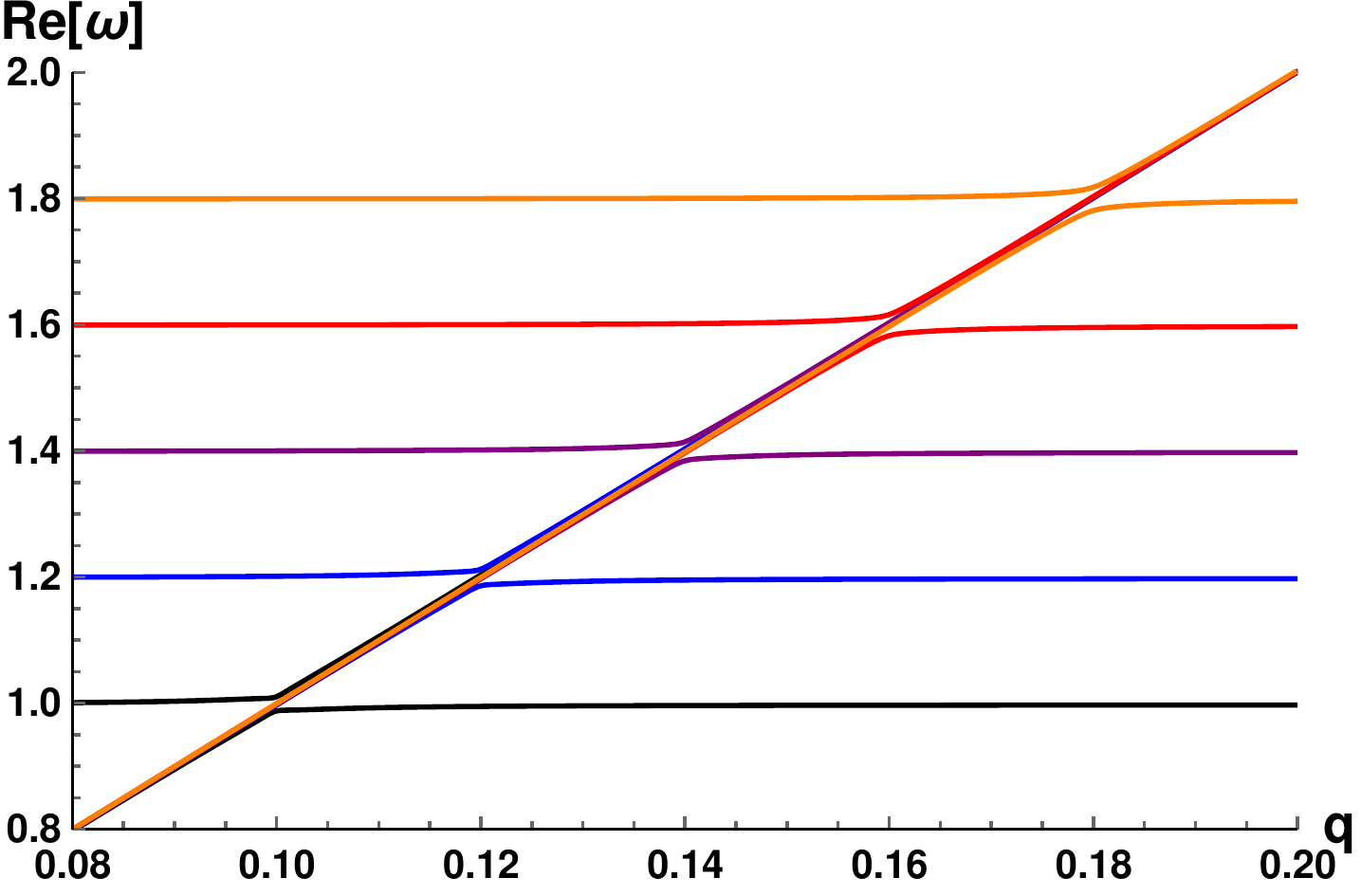}
\includegraphics[width=0.49 \linewidth]{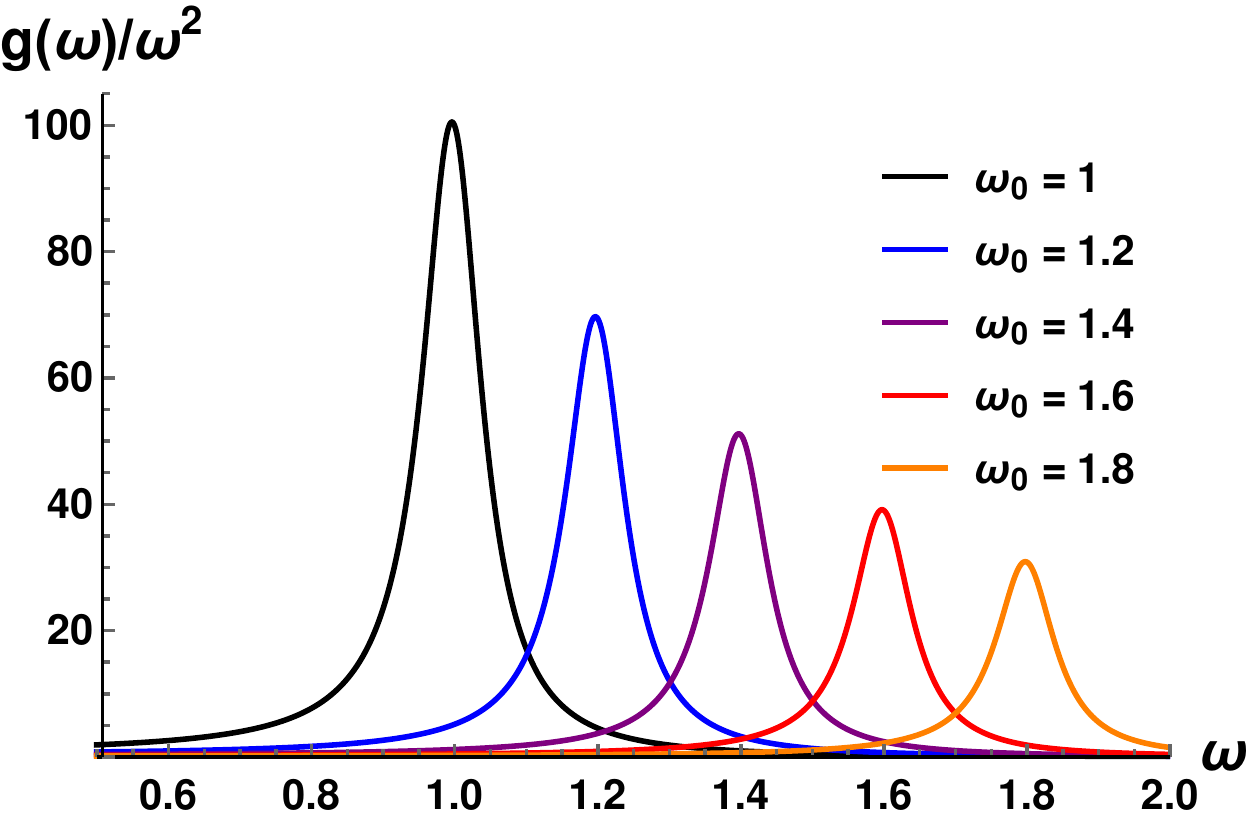}

\vspace{0.2cm}

\includegraphics[width=0.49 \linewidth]{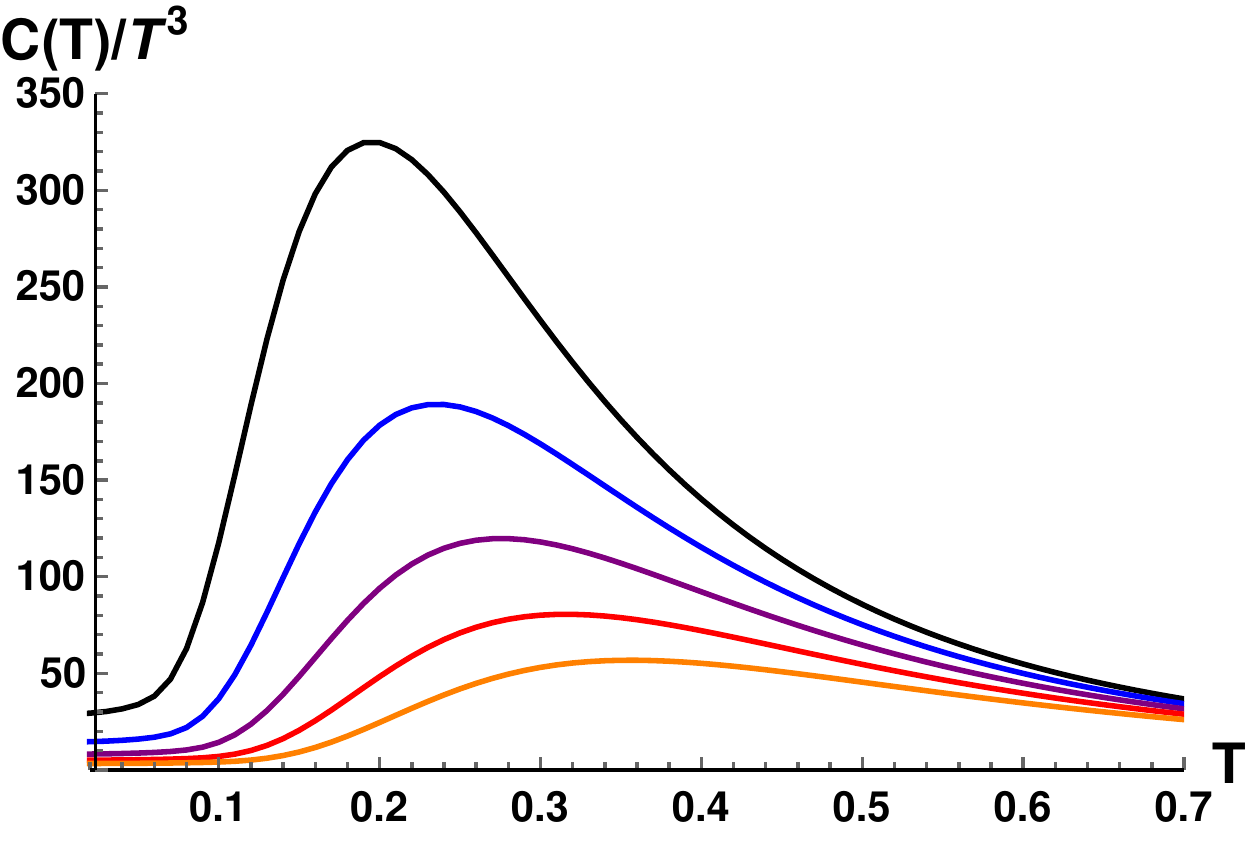}
\caption{\textbf{(Color online) Left: }The dispersion relation of the two branches of modes in the extended KK model. The different colors correspond to different values for the energy of the optical-like mode $\omega_0$. \textbf{Right: } The corresponding normalized VDOS. \textbf{Bottom: } The normalized specific heat.}
\label{panel3}
\end{figure}

\begin{figure}[h]
\centering
\includegraphics[width=0.49 \linewidth]{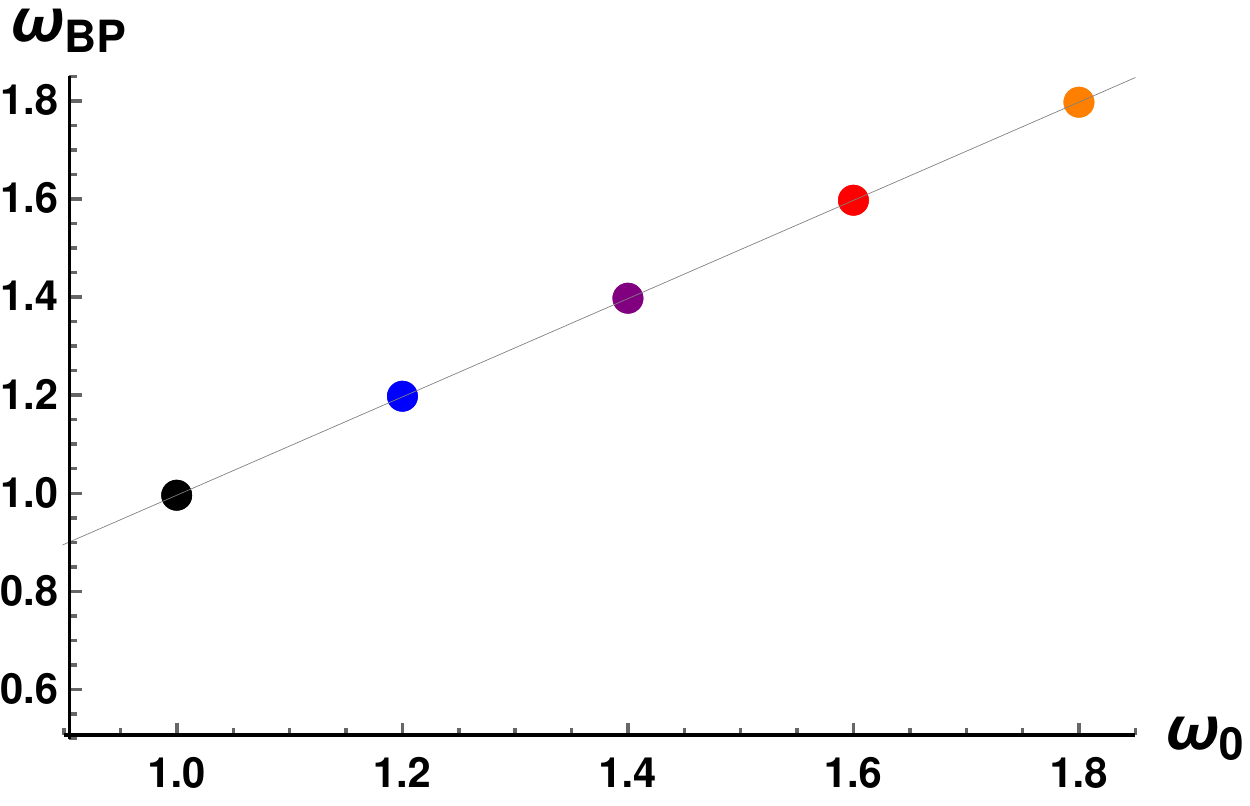}
\includegraphics[width=0.49 \linewidth]{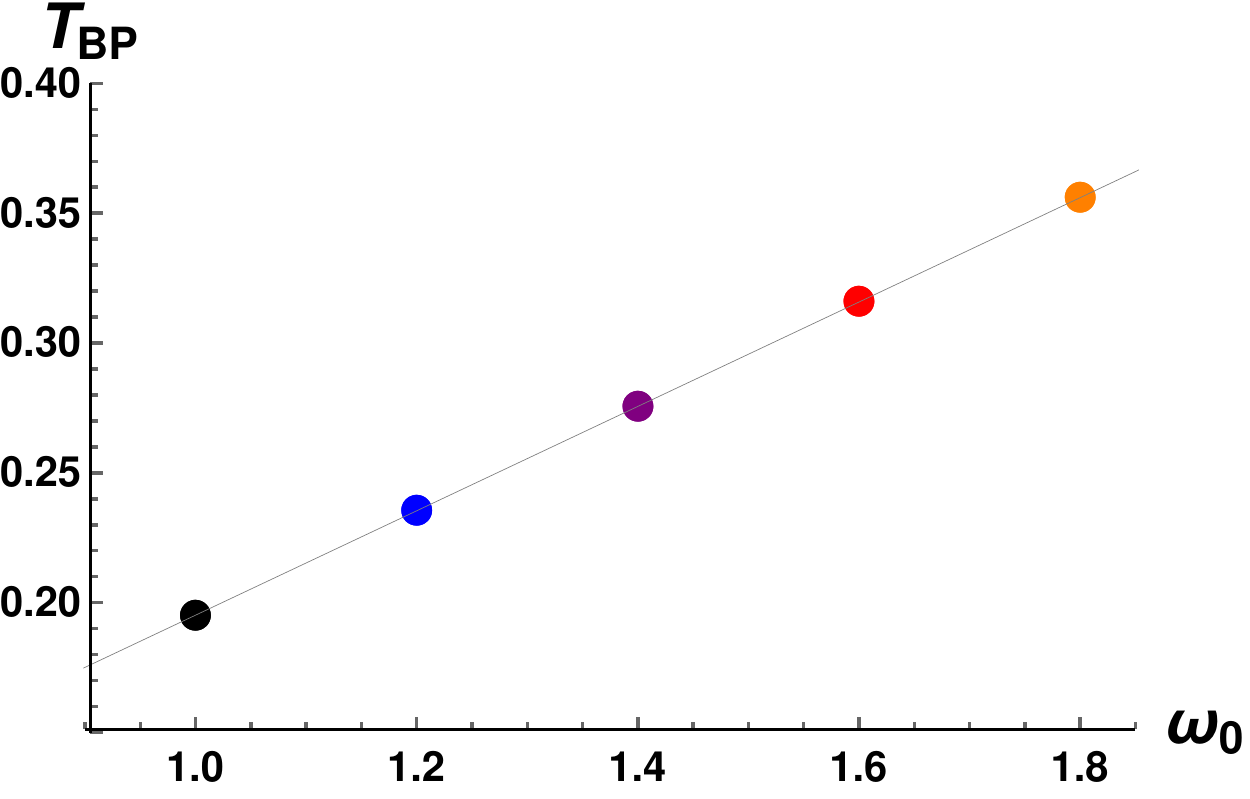}
\caption{(Color online) The evolution of the boson peak frequency and of the boson peak temperature as a function of the energy of the soft modes $\omega_0$. The linear correlation is evident as confirmed by empirically fitting the numerical data to a linear relation.}
\label{panel4}
\end{figure}

We can now study the behaviour as a function of the characteristic energy scale $\omega_0$, which corresponds to the energy (or energy gap) of the optical-like soft modes. Here we keep the density of defects $c$ constant, such that the energy separation $\Delta \omega$ between the two modes is approximately constant. The main results are presented in Fig.\ref{panel3}. We observe that the strength of the boson peak in both the VDOS and the specific heat becomes weaker upon increasing $\omega_0$. In other words, only soft optical-like modes, whose energy is not too large compared to the energy scale of the acoustic-like modes, contribute to the low energy glassy-like behaviour. Moreover, increasing the energy of the (no longer) soft modes, the boson peak moves towards higher energy. In particular, we notice a direct correlation between the position of the boson peak and the frequency $\omega_0$ of the soft optical-like modes, which is shown in Fig.\ref{panel4}. 

Both the boson peak frequency $\omega_{BP}$ and the boson peak temperature $T_{BP}$ display a very clear linear scaling in terms of the $\omega_0$ energy parameter. 
Furthermore, it is seen that the boson peak frequency is very close to the frequency at which the van Hove singularity occurs, i.e. the frequency at which  band-flattening, $d\omega/dq=0$, occurs.
This suggests that the origin of the boson peak in these materials is related to a van Hove singularity caused by band-flattening of the polariton-like spectrum, smeared by anharmonicity.

In conclusion, we derived an extended version of the Klinger-Kosevich model of soft mode dynamics \cite{KKmodel}, which crucially accounts for viscous damping of vibrational modes, to predict the vibrational anomalies experimentally observed in thermoelectric host-guest materials, such as the boson peak observed in the VDOS~\cite{doi:10.1002/adma.201706230}. The theory shows, semi-analytically, that the presence of rattlers, associated with soft-gapped quasi-localized modes, and the avoided crossing feature produced by their interactions with the acoustic-like phonons, are the fundamental processes leading to a "boson peak'' in the VDOS caused by a van Hove singularity from band-flattening of the polariton-like spectrum, smeared out by anharmonicity. Our results show that, upon increasing the density of guest atoms, the strength of the boson peak (BP) in the VDOS becomes larger and the BP moves in a power-law fashion towards lower frequencies. This result quantitatively establishes the idea that the VDOS of thermoelectrics can be tuned by the density of the guest defects, e.g. by the stoichiometry in tetrahedrites where this BP effect has been measured experimentally. Additionally, we observe a strong linear correlation between the position of the BP and the energy of the optical-like soft modes. This provides a further confirmation regarding the possible glassy-like effects induced by softly gapped degrees of freedom, like the soft optical phonons considered in \cite{baggioli2018soft}. 

This simple theoretical model successfully explains the peak in the VDOS observed experimentally in thermoelectric tetrahedrites~\cite{doi:10.1002/adma.201706230}. Moreover, together with recent experimental and theoretical results \cite{doi:10.1021/acs.jpcb.5b04240,PhysRevB.99.024301,Nakayama_2008,BaggioliPRL}, it opens up the way of realizing technologically relevant materials with crystal-like electronic behaviour and glass-like phononic behaviour, where the boson peak can be tuned by stoichiometry in order to minimize the thermal conductivity of the material. For example, in the thermoelectric tetrahedrite materials studied in ~\cite{doi:10.1002/adma.201706230} the stoichimetry of the Cu atoms is directly related to the parameter $c$ used in our model.
The presence of these features seems to be more universal and general than thought before and presumably tightly connected with anharmonic damping mechanisms and softly-gapped modes.\\

Our results could have immediate generalizations to the study of polaritonic systems displaying avoided crossing. Moreover, they suggest a possible fundamental role of softly-gapped vibrational modes for the onset of non-standard superconductivity.

\vspace{0.5cm}

{\centering \textbf{Acknowledgments}}\\[0.1cm]
We thank Giuseppe Carini, Tatsuya Mori, and Miguel Ramos for fruitful discussions and interesting comments about the topics considered in this manuscript. We thank particularly Josep-Lluis Tamarit for reading a preliminary version of the manuscript and for providing useful comments.
M.B. acknowledges the support of the Spanish MINECO’s “Centro de Excelencia Severo Ochoa” Programme under grant SEV-2012-0249.

\bibliographystyle{apsrev4-1}
\bibliography{KK}

\end{document}